\documentclass[floats,floatfix,showpacs,amssymb,prl,twocolumn,superscriptaddress,nofootinbib,reprint]{revtex4-1}

\usepackage{graphicx,epsf, epsfig, amssymb}
\usepackage{bm}
\usepackage{longtable}
\usepackage[usenames,dvipsnames]{xcolor}
\usepackage{color}
\usepackage[colorlinks=true,linktocpage=false,citecolor=blue,linkcolor=blue,urlcolor=blue]{hyperref}
\usepackage{amsfonts,amsmath,amssymb,mathrsfs}
\usepackage{multirow}
\usepackage[nolist,nohyperlinks]{acronym}
\usepackage{xspace}
\usepackage{booktabs,array}

\def\etal{et al.\xspace}
\def\bns{\ac{NS}-\ac{NS}\xspace}
\def\bbh{\ac{BH}-\ac{BH}\xspace}
\def\nsbh{\ac{NS}-\ac{BH}\xspace}
\def\sacra{{\ttfamily SACRA}\xspace}

\def\be{\begin{equation}}
\def\ee{\end{equation}}
\def\beq{\begin{eqnarray}}
\def\eeq{\end{eqnarray}}
\def\ben{\begin{enumerate}}
\def\een{\end{enumerate}}
\def\bi{\begin{itemize}}
\def\ei{\end{itemize}}

\newcommand{\fCut}{f_{\rm cut}}

\newcommand{\mBH}{M_{\rm BH}}
\newcommand{\mNS}{M_{\rm NS}}
\newcommand{\mSun}{M_\odot}

\newcommand{\rNS}{R_{\rm NS}}

\newcommand{\fRD}{f_{\rm RD}}

\newcommand{\fTide}{f_{\rm tide}}
\newcommand{\mTorus}{M_{\rm b, torus}}

\begin{document}

\title{Gravitational-wave cutoff frequencies of tidally disruptive\\
  neutron star-black hole binary mergers}

\author{Francesco Pannarale}
\email{francesco.pannarale@ligo.org}
\affiliation{School of Physics and Astronomy, Cardiff University, The
  Parade, Cardiff CF24 3AA, UK}

\author{Emanuele Berti}
\affiliation{Department of Physics and Astronomy, The University of
  Mississippi, University, MS 38677, USA}
\affiliation{CENTRA, Departamento de F\'isica, Instituto Superior
  T\'ecnico, Universidade de Lisboa, Avenida Rovisco Pais 1, 1049
  Lisboa, Portugal}

\author{Koutarou Kyutoku} 
\affiliation{Interdisciplinary Theoretical Science (iTHES) Research
  Group, RIKEN, Wako, Saitama 351-0198, Japan}

\author{Benjamin D. Lackey} 
\affiliation{Department of Physics, Princeton University, Princeton,
  NJ 08544, USA}
\affiliation{Department of Physics, Syracuse University, Syracuse, NY
  13244, USA}

\author{Masaru Shibata}
\affiliation{Yukawa Institute for Theoretical Physics, Kyoto
  University, Kyoto 606-8502, Japan}


\date{\today}

\begin{abstract}
  Tidal disruption has a dramatic impact on the outcome of neutron
  star-black hole mergers. The phenomenology of these systems can be
  divided in three classes: nondisruptive, mildly disruptive or
  disruptive. The cutoff frequency of the gravitational radiation
  produced during the merger (which is potentially measurable by
  interferometric detectors) is very different in each regime, and
  when the merger is disuptive it carries information on the neutron
  star equation of state.  Here we use semianalytical tools to derive
  a formula for the critical binary mass ratio $Q=\mBH/\mNS$ below
  which mergers are disruptive as a function of the stellar
  compactness $\mathcal{C}=\mNS/\rNS$ and the dimensionless black hole
  spin $\chi$.  We then employ a new gravitational waveform amplitude
  model, calibrated to $134$ general relativistic numerical
  simulations of binaries with black hole spin (anti-)aligned with the
  orbital angular momentum, to obtain a fit to the gravitational-wave
  cutoff frequency in the disruptive regime as a function of
  $\mathcal{C}$, $Q$ and $\chi$.  Our findings are important to build
  gravitational wave template banks, to determine whether neutron
  star-black hole mergers can emit electromagnetic radiation (thus
  helping multimessenger searches), and to improve event rate
  calculations for these systems.
\end{abstract}

\maketitle

\begin{acronym}
\acrodef{BH}[BH]{black hole}
\acrodef{EM}[EM]{electromagnetic}
\acrodef{EOS}[EOS]{equation of state}
\acrodef{GW}[GW]{gravitational-wave}
\acrodef{IMR}[IMR]{inspiral-merger-ringdown}
\acrodef{ISCO}[ISCO]{innermost stable circular orbit}
\acrodef{KAGRA}[KAGRA]{Kamioka Gravitational wave detector}
\acrodef{LIGO}[LIGO]{Laser Interferometer Gravitational-Wave Observatory}
\acrodef{NS}[NS]{neutron star}
\acrodef{PN}[PN]{post-Newtonian}
\acrodef{QNM}[QNM]{quasinormal mode}
\acrodef{SGRB}[SGRB]{short-hard gamma-ray burst}
\end{acronym}

\noindent{\bf{\em I. Introduction.}}~The merger of \acp{BH} and
\acp{NS} is one of the most violent events in the Universe.
Coalescing \nsbh systems are among the leading candidate sources for
upcoming interferometric \ac{GW} detectors such as the Advanced
\ac{LIGO} \cite{TheLIGOScientific:2014jea, 2010CQGra..27h4006H}, Virgo
\cite{TheVirgo:2014hva}, the \ac{KAGRA} \cite{Somiya:2011np,
  AsoKAGRA}, and \ac{LIGO}-India \cite{indigo}.  \ac{GW} observations
of \nsbh mergers may provide information on the \ac{NS}
\ac{EOS}~\cite{Read:2009yp} and on the underlying theory of
gravity~\cite{Berti:2015itd}. \nsbh binaries are also \ac{SGRB}
progenitor candidates~\cite{Paczynski:1991aq}.  If the \ac{NS} is
tidally disrupted during the merger, a hot disk with mass $\gtrsim
0.01\mSun$ may form around the spinning remnant \ac{BH}.  A scenario
where the \ac{BH}-disk system launches a relativistic jet by releasing
its gravitational energy via neutrino or \ac{EM} radiation on a time
scale $\lesssim 2$~s explains the duration, energetics, and estimated
event rates of \acp{SGRB}~\cite{Nakar:2007yr, Berger:2013jza}.  During
their merger, \nsbh binaries may also emit \ac{EM} radiation
isotropically --- as opposed to beamed \ac{SGRB} emission --- when
they eject unbound material.  This can be as massive as $\sim
0.1M_\odot$ and have subrelativistic velocities of $\sim
0.2$-$0.3c$~\cite{Kyutoku2015}, producing \ac{EM} counterparts in the
form of macronov\ae/kilonov\ae, powered by decay heat of unstable
$r$-process elements and by nonthermal radiation from electrons
accelerated at blast waves between the merger ejecta and the
interstellar medium~\cite{LiPaczynski1998, Kulkarni2005, Metzger2010,
  Nakar2011, TakamiH2014, Kisaka2015}.

The features of \ac{GW} emission from \nsbh binaries, as well as the
plausibility of these systems being central \ac{SGRB} engines and
sites for \ac{EM} radiation emission in general, depend crucially on
whether or not the \ac{NS} is tidally disrupted. Only numerical
simulations can assess this. Fortunately, with the enormous progress
made over the last decade, numerical relativity has provided a clear
picture of \nsbh \ac{GW} emission and shedded light on the processes
leading to disk formation and mass ejection.  Most of the \ac{GW}
emission occurs prior to the \ac{NS} tidal disruption, if this happens
at all, and before significant thermal effects take place.  Further,
magnetic fields appear to barely affect the \ac{GW} signal.  These
circumstances imply that an ideal fluid-dynamics treatment with a cold
\ac{EOS} is appropriate to simulate the dynamical regime of interest
for \ac{GW} signals.  A notable feature of these signals is that the
cutoff frequency at which their amplitude damps due to the \ac{NS}
tidal disruption depends on the \ac{NS} \ac{EOS}.  Hence, the cutoff
frequency encodes information on the \ac{EOS} itself, particularly
when this is stiff~\cite{Shibata:2009cn, Ferrari:2009bw,
  Kyutoku:2010zd, Kyutoku:2011vz}.

\begin{figure*}[!tb]
  \begin{tabular*}{\textwidth}{c@{\extracolsep{\fill}}c@{\extracolsep{\fill}}c}
    \includegraphics[width=0.33\textwidth,clip=true]{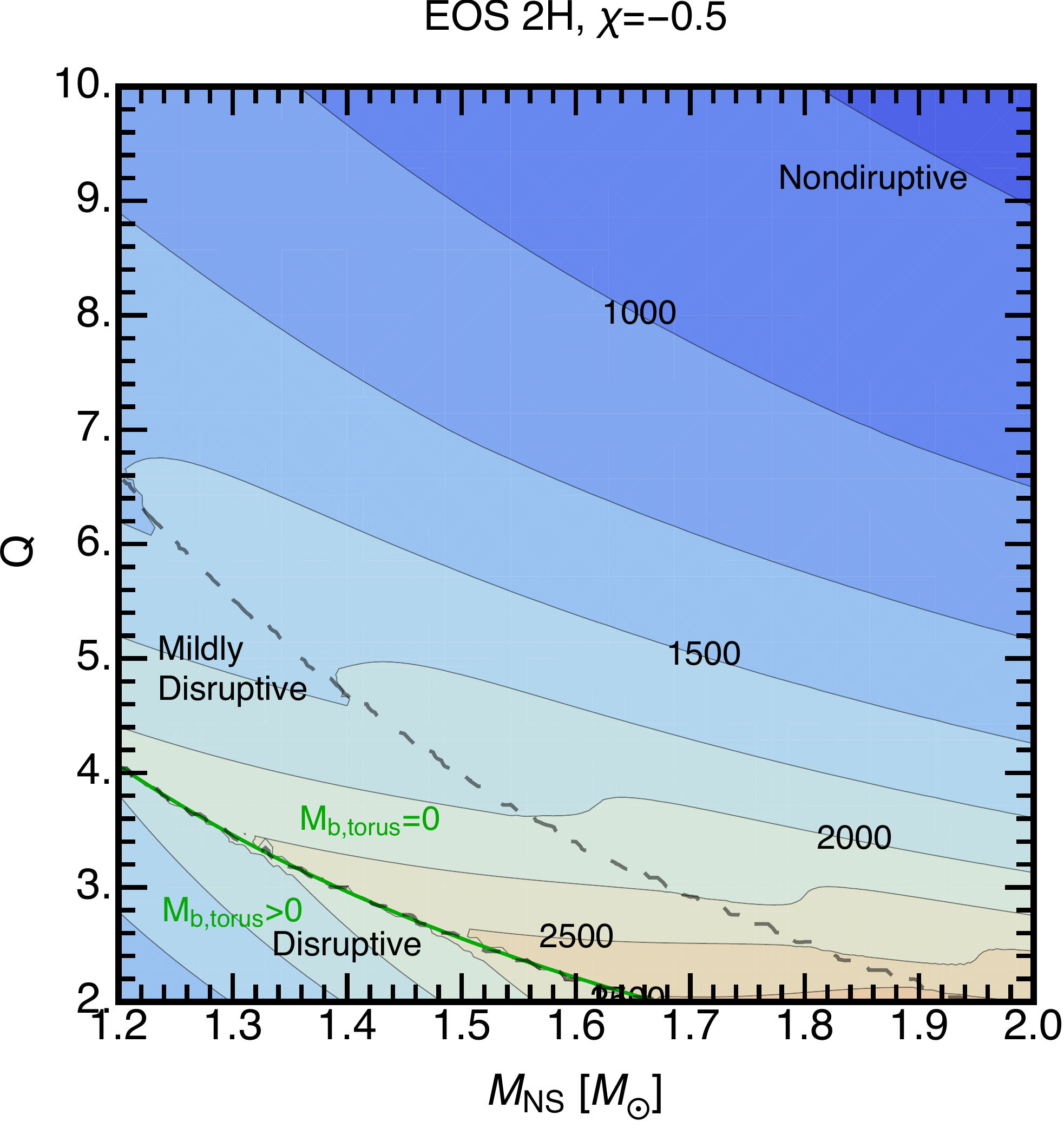}&
    \includegraphics[width=0.33\textwidth,clip=true]{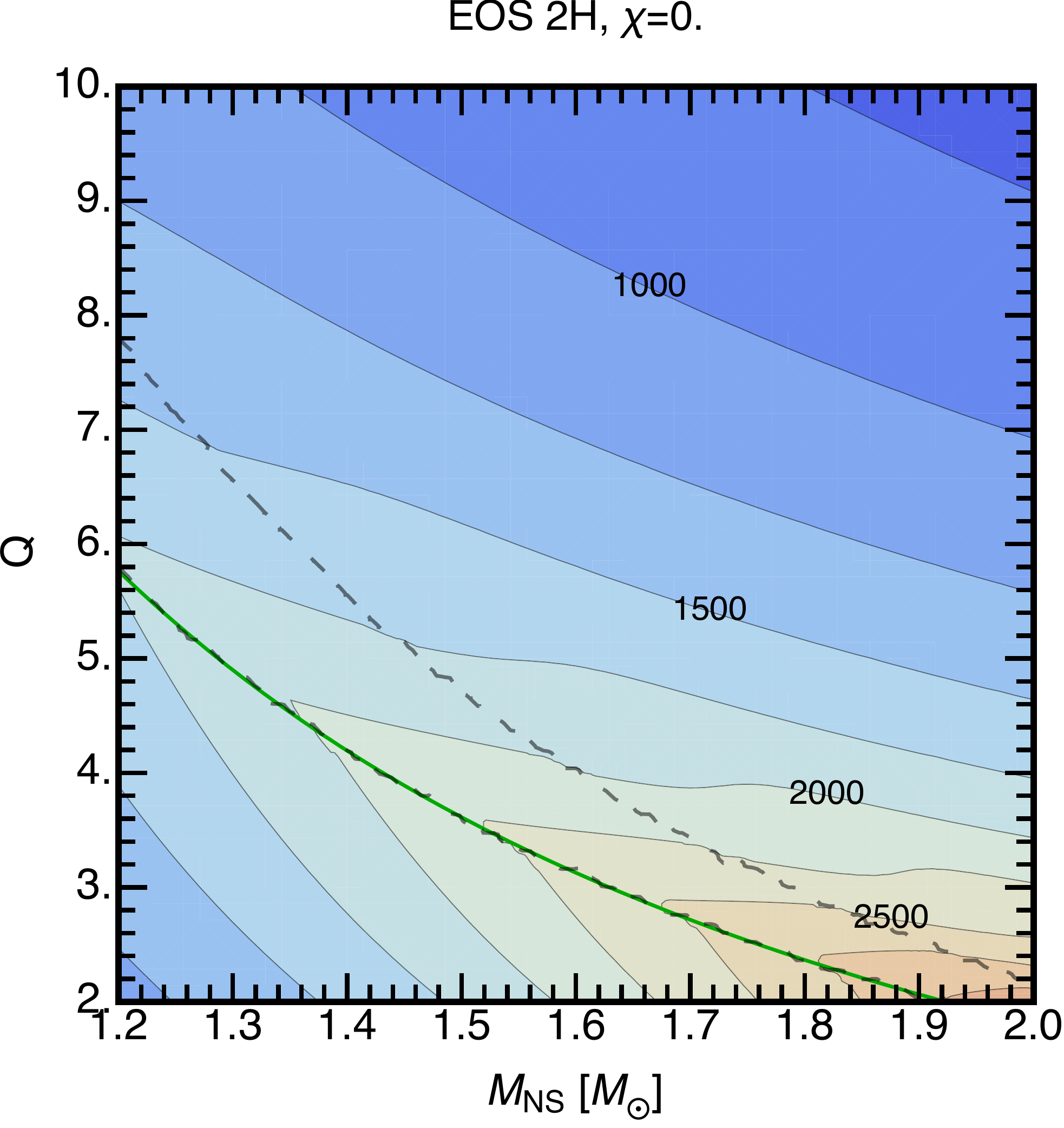}&
    \includegraphics[width=0.33\textwidth,clip=true]{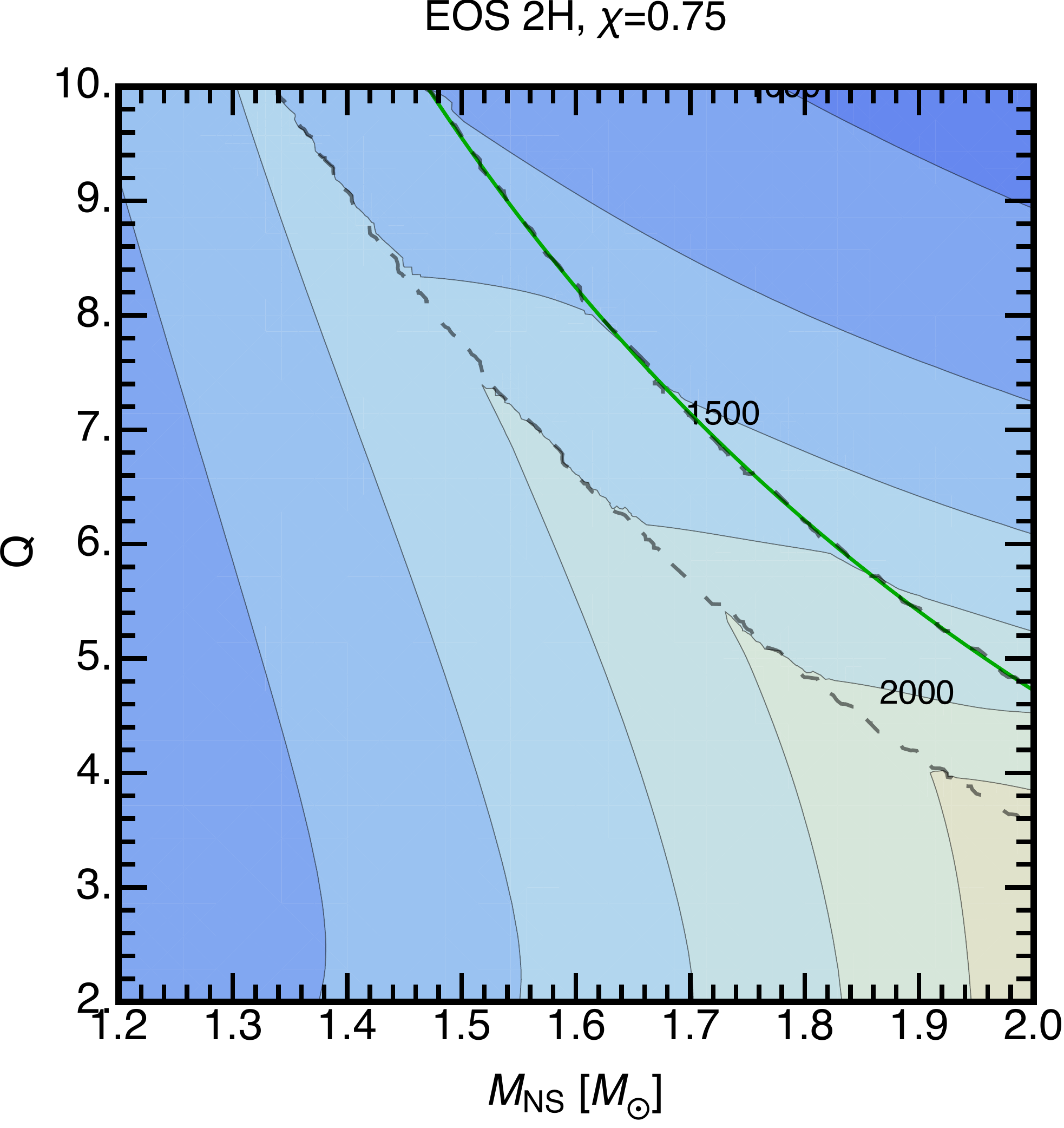}\\
    \includegraphics[width=0.33\textwidth,clip=true]{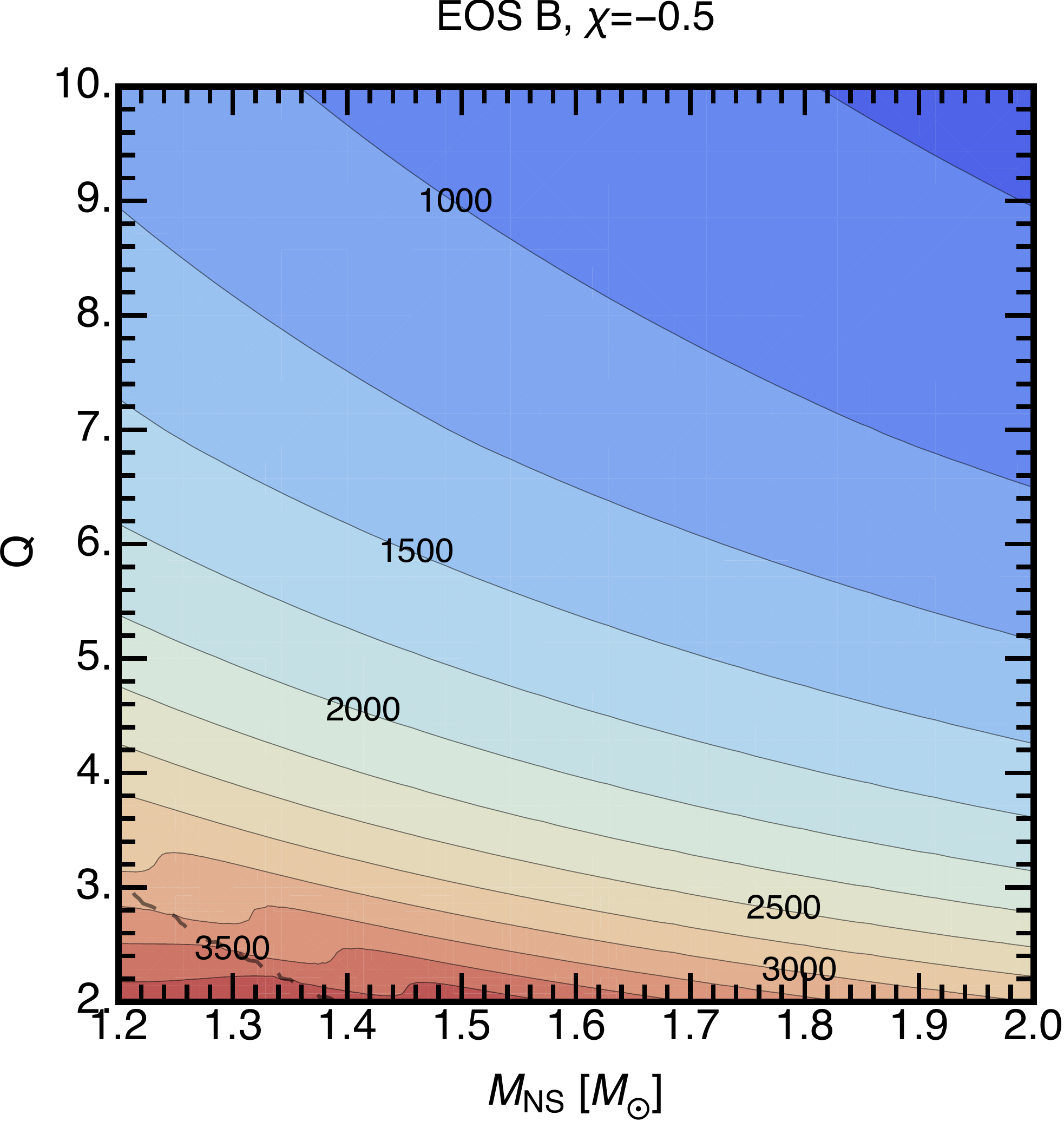}&
    \includegraphics[width=0.33\textwidth,clip=true]{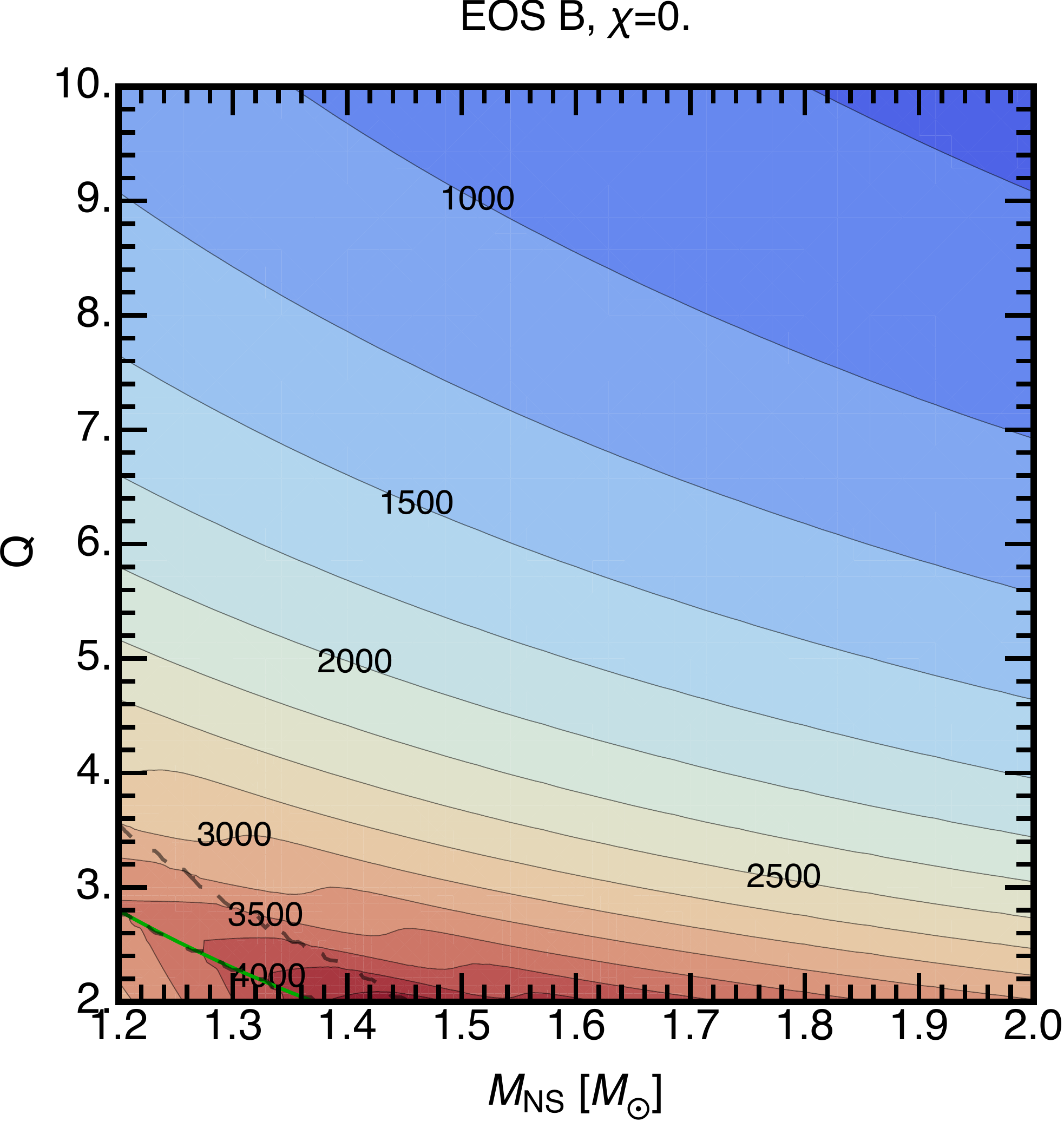}&
    \includegraphics[width=0.33\textwidth,clip=true]{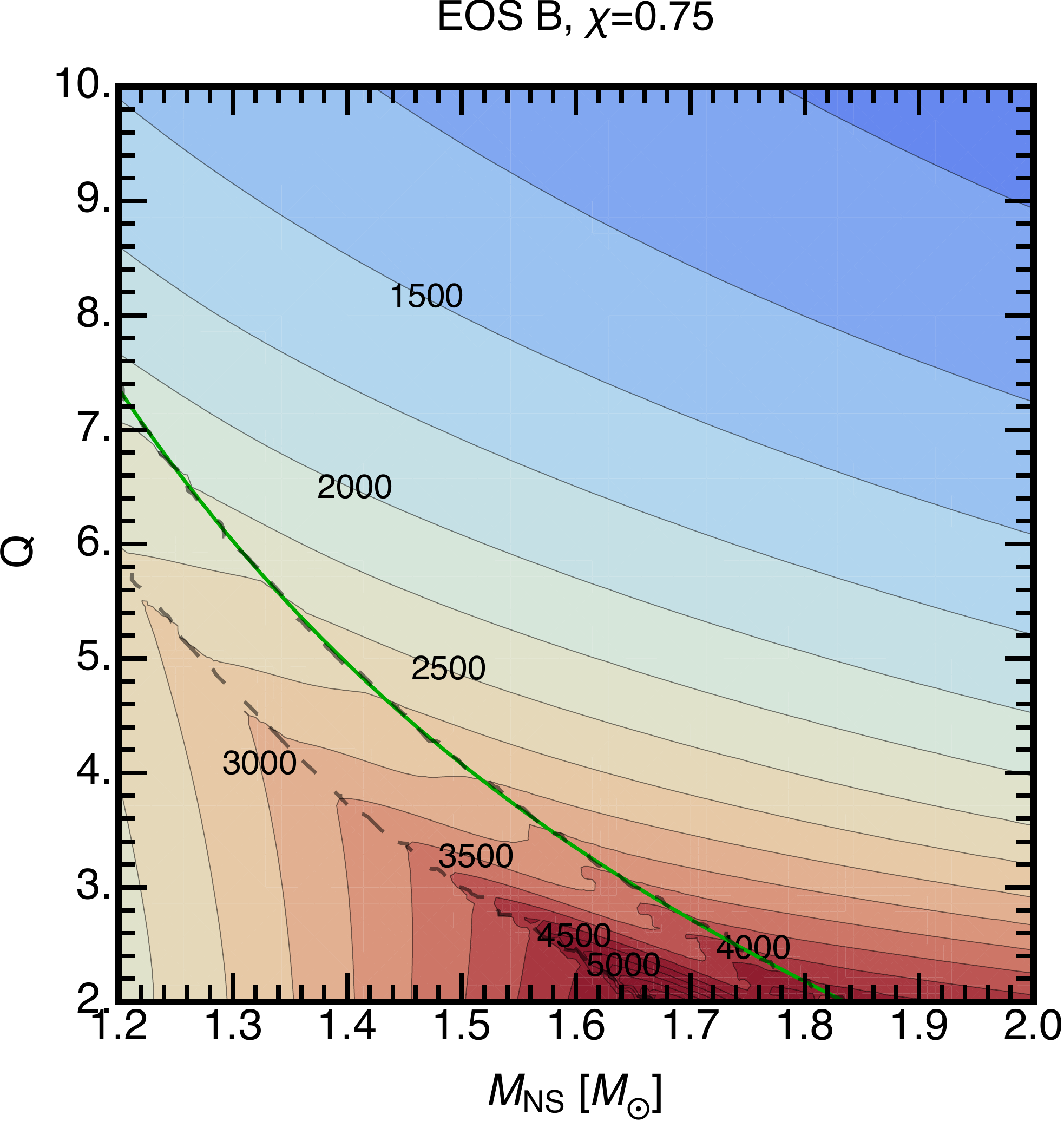}\\
  \end{tabular*}
  \caption{The cutoff frequency $\fCut$, as defined in
    Eq.\,(\ref{eq:fCut}), computed with our \nsbh \ac{GW} amplitude
    model~\cite{PRDprep}.  Each panel label specifies the \ac{NS}
    \ac{EOS} and the \ac{BH} spin parameter $\chi$ used.  The contour
    lines report $\fCut$ in Hz and have a spacing of $250\,$Hz.  The
    thick, green, continuous line is the location where the mass of
    the torus remnant $\mTorus$ vanishes.  The two dashed lines in
    each panel divide the plane in three regions: a top-right region
    in which \nsbh coalescences are nondisruptive ($\fTide\geq\fRD$
    and $\mTorus=0$), a bottom-left one in which they are disruptive
    ($\fTide<\fRD$ and $\mTorus>0$), and a middle region in which
    mildly disruptive coalescences occur ($\fTide<\fRD$ and
    $\mTorus=0$, or $\fTide\geq\fRD$ and
    $\mTorus>0$).  \label{fig:fCutContours_chiall}}
\end{figure*}

Numerical simulations of compact binary mergers are still very
resource intensive, which makes semianalytical models very valuable.
Simulations and models are most advanced for \bbh systems.  Waveform
models belong to two main classes: Fourier-domain phenomenological
\ac{IMR} models based on a \ac{PN} description of the early inspiral
stage (``PhenomA''~\cite{Ajith:2007qp, Ajith:2007kx, Ajith:2007xh},
``PhenomB''~\cite{Ajith:2009bn}, ``PhenomC''~\cite{Santamaria:2010yb},
and ``PhenomP''~\cite{PhenomP}) and effective-one-body models tuned to
\bbh simulations~\cite{Taracchini:2012ig, Damour:2012ky,
  Damour:2010zb, Ohme:2011zm, Pan2014, Taracchini2014, Szilagyi2015}.
\nsbh waveform models are far less developed~\cite{Lackey2014, PaperI,
  PRDprep}, because long, accurate simulations are particularly hard
to achieve and because the parameter space is larger. The outcome and
aftermath of \bns and \nsbh binary simulations (as opposed to \bbh
ones) depend on several assumptions on currently underconstrained
physics, such as the \ac{NS} \ac{EOS}, magnetic field configurations,
and neutrino emission. The relatively high mass ratios in \nsbh
systems cause both analytical and numerical complications: the
convergence of the \ac{PN} approximation is slower than for \bns
systems~\cite{Berti:2007cd,Buonanno:2009zt}, initial data are hard to
construct~\cite{Kyutoku:2014yba} and the simulations must track very
different dynamical time scales. In fact, \bbh \ac{GW} templates are
commonly used in \nsbh merger searches.  Tidal disruption affects both
the \ac{GW} and \ac{EM} emission of \nsbh binaries. For all these
reasons, better models can directly improve \ac{GW} template banks,
determine whether \nsbh mergers can power \acp{SGRB} and emit \ac{EM}
radiation in general (thus helping multimessenger searches), and
improve event rate calculations for these systems.

\noindent{\bf{\em II. Simulations and model.}} At least two papers
attempted a phenomenological description of the \acp{GW} emitted by
\nsbh binaries.  Lackey \etal \cite{Lackey2014} developed an analytic
representation of the \ac{IMR} waveform calibrated to $134$ numerical
waveforms produced by the \sacra code~\cite{SACRA} with the goal of
assessing the measurability of the \ac{NS} tidal deformability.  The
subset of simulations for non-spinning \acp{BH} was used by Pannarale
\etal~\cite{PaperI} to build a frequency domain phenomenological
waveform \emph{amplitude} model which was, at heart, a ``distortion''
of the PhenomC \bbh model.  This model relied on the fit of
numerical-relativity results presented in~\cite{Foucart:2012nc} to
compute the remnant torus mass $\mTorus$.  It paid particular
attention to the accuracy at high frequencies -- where the
\ac{EOS}-related phenomenology takes place -- and to the determination
of the \ac{GW} cutoff frequency.  In units in which the total mass of
the system is set to unity, let $h(f)$ be the Fourier transform of the
\ac{GW} signal, and $f_{\rm Max}$ the frequency at which $f^2h(f)$ is
maximum.  We define $\fCut$ ($>f_{\rm Max}$) as the frequency at which
the dimensionless amplitude drops by one $e$-fold:
\be
\label{eq:fCut}
e \fCut h(\fCut) = f_{\rm Max}h(f_{\rm Max})\,.%
\ee
Figure \ref{fig:fCutContours_chiall} displays $\fCut$ for a sample of
\ac{BH} spin parameters and for two piecewise polytropic \ac{EOS}
models (2H and B), chosen because they yield low- and high-compactness
\acp{NS}, respectively (see~\cite{PRDprep}).  This cutoff frequency is
important to construct \ac{GW} template banks for \nsbh searches:
targeting \nsbh binaries with \bbh templates terminated at a frequency
$f_{\rm term}<\fCut$ results in a signal-to-noise ratio loss; on the
contrary, using $f_{\rm term}>\fCut$ may penalize the template by
degrading its chi-square test performance, as it lacks matter effects.

The companion paper~\cite{PRDprep} extends the work of~\cite{PaperI}
to \nsbh systems with a non-precessing, spinning \ac{BH}, using the
full set of $134$ hybrid waveforms constructed in~\cite{Lackey2014}.
These are based on simulations in which the \ac{NS} matter at zero
temperature is modeled via piecewise polytropic \acp{EOS} that mimic
nuclear-theory-based \acp{EOS} with a small number of
parameters~\cite{Read:2008iy}.  The binary mass ratio takes the values
$Q\in \{2$, $3$, $4$, $5\}$ and the \ac{BH} dimensionless spin
parameter $\chi\in \{-0.5$, $0$, $0.25$, $0.5$, $0.75\}$.  This
parameter space coverage allows our model to produce the most accurate
prediction of cutoff frequencies for \nsbh \ac{GW} signals, with
relative errors below $10$\%, well below the errors one would obtain
using either \bbh models or the \nsbh model of~\cite{Lackey2014}.  The
new \nsbh \ac{GW} amplitude model used here, and detailed in
Ref.~\cite{PRDprep}, is adapted to the three possible fates of the
binary (see also Fig.\,\ref{fig:fCutContours_chiall}): (1) {\em
  nondisruptive}: the \ac{GW} frequency at the onset of tidal
disruption $\fTide\geq \fRD$ (where $\fRD$ is the \ac{BH} remnant
dominant ringdown frequency, calculated as in~\cite{Pannarale2012,
  Pannarale2014}, and $\fTide$ is the frequency at the onset of
mass-shedding, determined as in~\cite{Foucart:2012nc}), and $\mTorus$,
computed as in~\cite{Foucart:2012nc}, vanishes; (2) {\em
  disruptive}:~$\fTide<\fRD$ and $\mTorus>0$; or (3) {\em mildly
  disruptive}: either $\fTide<\fRD$ and $\mTorus=0$, or
$\fTide\geq\fRD$ and $\mTorus>0$.  The \ac{BH} ringdown does not
contribute to the \ac{GW} emission of disruptive mergers.  Ringdown
radiation appears in mildly disruptive mergers, and looks similar to
\ac{BH}-\ac{BH} mergers in nondisruptive cases.  Disruptive mergers
thus differ the most from \ac{BH}-\ac{BH} mergers, precisely because
tidal effects are maximal. In these events, $\fCut$ and the \ac{NS}
\ac{EOS} have a strong link.  To encompass all \nsbh binaries with a
\ac{GW} signal deviating from the \bbh case, one must consider
disruptive and mildly-disruptive mergers, i.e.~binaries above the top
dashed curves in Fig.\,\ref{fig:fCutContours_chiall} must be
discarded.  The remaining set of binaries includes the $\mTorus=0$
surface, and thus all possible \ac{EM} sources (within the
approximations of the model).

\begin{figure}[!tb]
  \includegraphics[width=\columnwidth,clip=true]{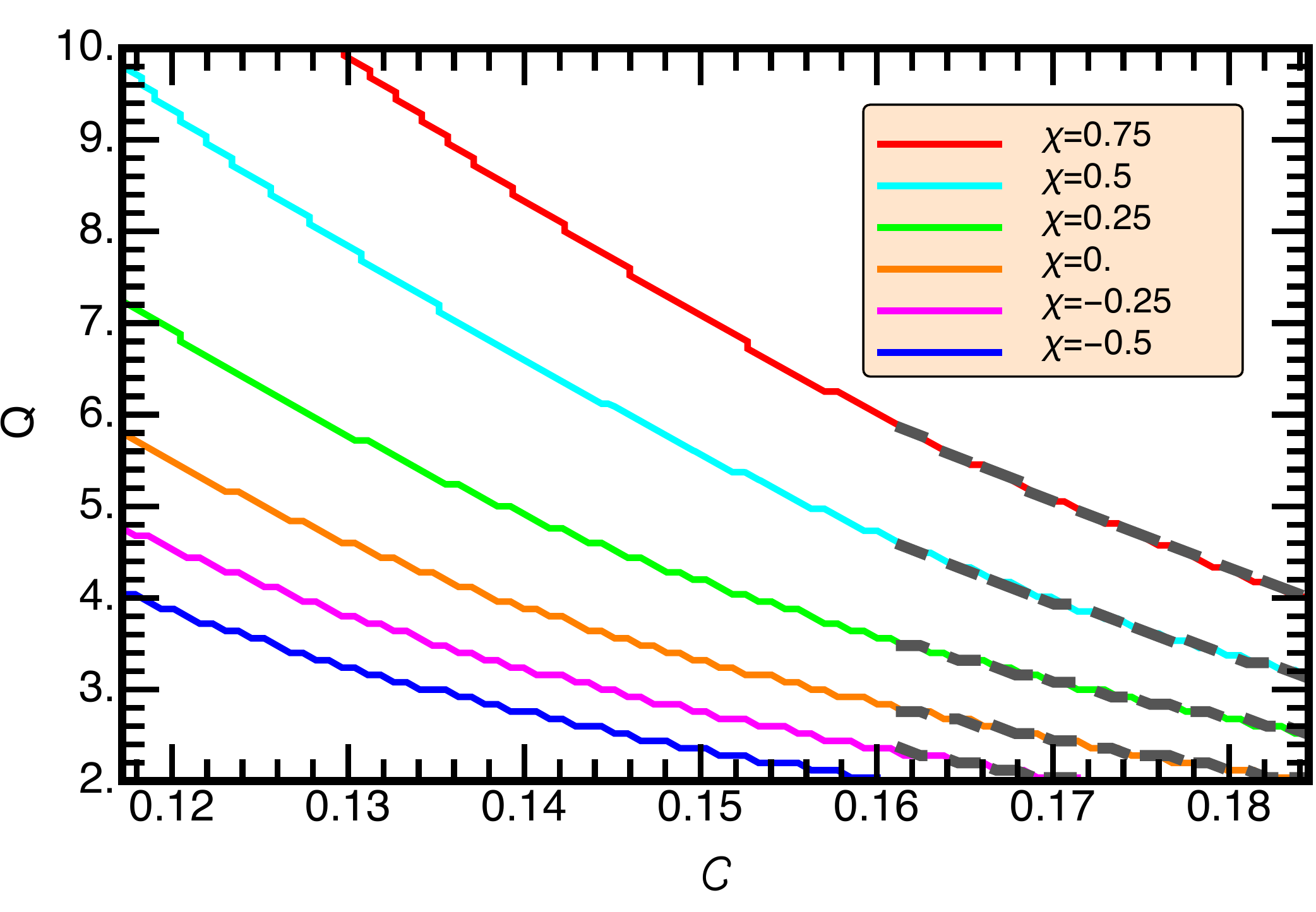}
  \caption{Boundaries separating disruptive mergers with $\fTide<\fRD$
    and $\mTorus>0$ (below each curve) from mildly disruptive and
    nondisruptive mergers (above each curve) for specific \ac{BH} spin
    values indicated in the legend from top to bottom.
    Continuous (dashed gray) lines refer to \ac{EOS} 2H
    (B).\label{fig:DMDboundaries}}
\end{figure}

\begin{table*}[!tb]
  \caption{\label{tab:fit_coeffs} Values of the coefficients of the fits in Eqs.\,(\ref{eq:DMDfit})-(\ref{eq:fCutFit}).  The number below each coefficient symbol must be multiplied by the corresponding power of ten in square brackets.  The $f_{ijk}$'s are expressed in units of $G=c={\rm total~mass}=1$.}
  \resizebox{\textwidth}{!}{%
    \begin{tabular}{c@{\hspace{0.3cm}}c@{\hspace{0.3cm}}c@{\hspace{0.3cm}}c@{\hspace{0.3cm}}c@{\hspace{0.3cm}}c@{\hspace{0.3cm}}c@{\hspace{0.3cm}}c@{\hspace{0.3cm}}c@{\hspace{0.3cm}}c@{\hspace{0.3cm}}}
      \toprule[1.pt]
      \toprule[1.pt]
      \addlinespace[0.2em]
      $a_{00}$ $[10^{1}]$ & $a_{10}$ $[10^{2}]$ & $a_{01}$ $[10^{1}]$ & $a_{20}$ $[10^{3}]$ & $a_{11}$ $[10^{2}]$ & $a_{02}$ $[10^{1}]$ & $a_{30}$ $[10^{3}]$ & $a_{21}$ $[10^{2}]$ & $a_{12}$ $[10^{1}]$ & $a_{03}$ $[10^{-1}]$ \\
      $4.59676$          & $-6.68812$         & $2.78668$          & $3.56791$           & $-2.79252$         &  $1.07053$         & $-6.69647$         & $7.55858$          & $-5.51855$         & $4.01679$ \\
      \addlinespace[0.2em]
      \midrule[1.pt]
      \addlinespace[0.2em]
      & $a_{0}$ $[10^{1}]$ & $a_{1}$ $[10^{1}]$ & $a_{2}$ $[10^{1}]$ & $a_{3}$ $[10^{1}]$ & $b_{0}$ $[10^{1}]$ & $b_{1}$ $[10^{0}]$ & $b_{2}$ $[10^{1}]$ & $b_{3}$ $[10^{1}]$ & \\
      & $5.52167$         & $4.05338$         & $3.09804$          & $-6.90163$       & $-1.68616$         & $-2.87849$        & $-1.82097$        & $1.36910$         & \\
      \addlinespace[0.2em]
      \midrule[1.pt]
      \addlinespace[0.2em]
      $f_{000}$ $[10^{-1}]$ & $f_{100}$  $[10^{0}]$ & $f_{010}$ $[10^{-2}]$ & $f_{001}$ $[10^{-2}]$ & $f_{200}$ $[10^{1}]$ & $f_{020}$ $[10^{-3}]$ & $f_{002}$ $[10^{-2}]$ & $f_{110}$ $[10^{-1}]$ & $f_{101}$ $[10^{-1}]$ & $f_{011}$  $[10^{-4}]$ \\
      $1.38051$            & $-2.36698$           & $-3.07791$          & $3.06474$            & $1.19668$           & $1.81262$            & $4.31813$            & $2.89424$            & $-1.61434$          & $9.30676$ \\
      \colrule
      \addlinespace[0.2em]
      $f_{300}$ $[10^{1}]$ & $f_{030}$ $[10^{-5}]$ & $f_{003}$ $[10^{-3}]$ & $f_{210}$ $[10^{-1}]$ & $f_{120}$ $[10^{-3}]$ & $f_{201}$ $[10^{-1}]$ & $f_{102}$ $[10^{-1}]$ & $f_{021}$ $[10^{-4}]$ & $f_{012}$ $[10^{-3}]$ & $f_{111}$ $[10^{-2}]$ \\
      $-1.46271$          & $-6.89872$           & $-2.29830$          & $2.73922$            & $-4.69093$           & $1.75728$            & $-2.04964$           & $5.52098$            & $-5.79629$          & $-9.09280$ \\
      \bottomrule[1.pt]
      \bottomrule[1.pt]
    \end{tabular}
  }
\end{table*}

\noindent{\bf{\em III. Predicting the fate of \nsbh mergers.}}~The two
dashed curves in the panels of Fig.\,\ref{fig:fCutContours_chiall}
separate disruptive mergers (bottom-left region), non-disruptive
mergers (top-right region), and mildly disruptive mergers (region in
between the two lines).  The green continuous line marks the locus of
binaries for which $\mTorus$ goes to zero.  Here we construct simple
formulas to quickly determine the fate of a \nsbh coalescence.  The
contours in Fig.\,\ref{fig:fCutContours_chiall} that separate \nsbh
binaries with a disruptive fate from those with a mildly disruptive or
nondisruptive fate may be fitted in several ways as a function of the
binary physical parameters.  We find it best to fit the critical mass
ratio $Q_{\rm D}$ \emph{below} which mergers are classified as
disruptive via a function of the form $Q_{\rm D}=Q_{\rm
  D}(\mathcal{C},\chi)$.  Fitting in terms of the \ac{NS} compactness
$\mathcal{C}$, rather than fixing an \ac{EOS} and fitting in terms of
the \ac{NS} mass, allows us to use at the same time data produced with
the two extreme \acp{EOS} B and 2H. This is evident in
Fig.\,\ref{fig:DMDboundaries}, where the boundaries between disruptive
and mildly disruptive or nondisruptive binaries are shown in the
$\mathcal{C}Q$-plane for initial \ac{BH} spin parameter values
$\chi\in \{-0.5, -0.25, 0., 0.25, 0.5, 0.75\}$: results for the 2H
\ac{EOS} (continuous) and the B \ac{EOS} (dashed) overlay.  This is
due to the fact that the criterion to determine whether a binary is
disruptive or not depends on $\mTorus$, and that the state-of-the-art
model for $\mTorus$~\cite{Foucart:2012nc} depends only on
$\mathcal{C}$ (i.e., it does not include higher-order,
\ac{EOS}-dependent effects).  Quite independently of the \ac{EOS}, the
merger of an \nsbh binary will be disruptive whenever $Q<Q_{\rm
  D}(\mathcal{C}, \chi)$, where the threshold is well fitted by
\begin{align}%
  \label{eq:DMDfit}
  Q_{\rm D} &= \sum_{i,j=0}^3 a_{ij} \mathcal{C}^i\chi^j\,, \quad
  i+j\leq 3\,,
\end{align}%
with coefficients reported in the first row of Table
\ref{tab:fit_coeffs}.  The relative errors between the data in
Fig.\,\ref{fig:DMDboundaries} and our fit are below $\sim
4\%$~\cite{PRDprep}.  Similarly, the relation
\begin{align}%
  \label{eq:NDMDfit}
  Q_{\rm ND} &= \left(\sum_{i=0}^{3} a_i\chi^i\right) \exp
  \left[\left(\sum_{j=0}^{3} b_j\chi^j\right)\mathcal{C}\right]\,,
\end{align}%
with coefficient values listed in the second row of Table
\ref{tab:fit_coeffs}, fits the boundary between nondisruptive and
mildly disruptive mergers so that \nsbh systems with $Q<Q_{\rm
  ND}(\mathcal{C}, \chi)$ are either disruptive or mildly disruptive.
The maximum relative error between the data and the fit is $\sim 8$\%,
and it is below $4.5$\% for $95$\% of the data points.

\noindent{\bf{\em IV. Cutoff frequency of disruptive
    mergers.}}~Equation~(\ref{eq:DMDfit}) allows us to determine when
a binary is disruptive according to our classification, i.e.~the
mass-shedding happens early enough during the evolution for the merger
to produce a remnant disk mass and the \ac{GW} emission deviates
significantly from the \bbh case at high frequencies.  We now wish to
provide a simple formula to compute the cutoff frequency of the
\ac{GW} amplitude $\fCut$ for disruptive mergers, as it carries
information on the nuclear \ac{EOS} and it may be valuable in building
better \ac{GW} template banks.  To this end we consider \ac{EOS} 2H,
generate a set of $10^4$ random disruptive mergers, compute $\fCut$
for each \nsbh binary, and finally fit the resulting data.  Disruptive
mergers are selected as follows: we randomly sample parameters in the
ranges $\mNS/M_\odot\in[1.2,2.83]$, $Q\in[2,10]$,
$\chi\in[-0.5,0.75]$; we verify whether the sampled point corresponds
to a disruptive binary, i.e.~whether $\fTide<\fRD$ and $\mTorus>0$; we
keep the point if it does; and we repeat the whole process until we
have the desired $10^4$ points.  While the maximum \ac{NS} mass for
the 2H \ac{EOS} is $\sim 2.83M_\odot$, the maximum \ac{NS} mass we
obtain for the sample of disruptive \nsbh mergers is $\sim
2.28M_\odot$. The resulting mass interval $\mNS/M_\odot\in[1.2,2.28]$
corresponds to a compactness interval of $0.117 \leq \mathcal{C}\leq
0.221$.  We then fit the data set with the function
\begin{align}%
  \label{eq:fCutFit}
  \fCut &= \sum_{i,j,k=0}^3 f_{ijk}
  \mathcal{C}^i Q^j \chi^k\,,
\quad i+j+k\leq 3\,.
\end{align}%
The resulting $f_{ijk}$ values are reported in Table
\ref{tab:fit_coeffs}.  The relative errors between this fit and the
data are typically below $1$\%: the relative error for $68$\%, $95$\%,
and $99.7$\% of the points is $0.47$\%, $1.5$\%, and $4.9$\%,
respectively.

As a consistency check, we draw a separate sample of $10^4$ disruptive
mergers, and, for each binary, compute $\fCut$ and the relative error
yielded by Eq.\,(\ref{eq:fCutFit}).  This time we use \ac{EOS} B,
which has a maximum \ac{NS} mass of $\sim 2M_\odot$.  The compactness
now ranges from $0.161$ to $0.225$.  Quite remarkably, the maximum
relative error is $2.2\%$; the relative error of $97.6$\% of the
points is below the percent level.  This check implies that
Eq.\,(\ref{eq:fCutFit}) is to a good
approximation~\ac{EOS}-independent, at least within the parameter
space where our model was calibrated and applied.

\noindent{\bf{\em V. Summary and discussion.}}~In this Letter we used
a recently developed semianalytical \ac{GW} amplitude model for \nsbh
mergers~\cite{PRDprep} to construct simple fits for:
(i) the critical binary mass ratio $Q_{\rm D}$ below which the merger
is disruptive [Eq.~\eqref{eq:DMDfit}] and its \ac{GW} emission
deviates significantly from a \bbh-like behavior;
(ii) the critical binary mass ratio $Q_{\rm ND}$ below which the
merger is either disruptive or mildly disruptive
[Eq.~\eqref{eq:NDMDfit}].  This can be viewed as a necessary but not
sufficient condition to generate an \ac{EM} counterpart, and may thus
be used to determine which binaries are plausible targets for
multimessenger searches targeting \acp{GW} and \ac{EM}/neutrino
emission;
(iii) the cutoff frequency $\fCut$ for disruptive mergers as a
function of the initial binary parameters $\mathcal{C}$, $Q$ and
$\chi$ [Eq.~\eqref{eq:fCutFit}]. This can be used to maximize the
recovered signal-to-noise ratio and chi-square test performance of
\bbh templates in \nsbh searches. The cutoff frequency can also be
used to constrain the \ac{NS} \ac{EOS}~\cite{Vallisneri:1999nq,
  Ferrari09, Ferrari:2009bw, Taniguchi:2008a, Shibata:2009cn,
  Kyutoku:2011vz, Kyutoku:2010zd}. Our fit suggests that measurements
of $\mathcal{C}$, $Q$ or $\chi$ from the inspiral radiation (see
e.g.~\cite{O'Shaughnessy:2014dka,Hannam:2013uu}) could improve the
resulting constraints on the \ac{EOS}.

The non-negligible eccentricity of the initial data used for the \nsbh
simulations underlying our model, the limited duration and finite
numerical resolution of the simulations, and the fitting errors all
limit the accuracy of $\fCut$ to a few percent, and therefore
introduce systematic errors.  These errors are expected to increase as
$Q\to Q_{\rm D}$, and when our fits are extrapolated beyond the region
where the model and fits were tuned. More importantly, it is necessary
to extend our model to precessing binaries~\cite{Kawaguchi:2015bwa}.
We plan to address these issues in future work.

\noindent
{\bf \em Acknowledgements.}
This work was supported by STFC grant No.~ST/L000342/1, by the
Japanese Grant-in-Aid for Scientific Research (21340051, 24244028),
and by the Grant-in-Aid for Scientific Research on Innovative Area
(20105004).  E.B.~is supported by NSF CAREER Grant PHY-1055103 and by
FCT contract IF/00797/2014/CP1214/CT0012 under the IF2014 Programme.
K.K.~ is supported by the RIKEN iTHES project.  B.L.~was supported by
NSF grants PHY-1305682, PHY-1205835, and AST-1333142.  F.P.~wishes to
thank Alex Nielsen, Alessandra Buonanno, Stephen Fairhurst, Tanja
Hinderer, and Bangalore Sathyaprakash for interesting discussions
throughout the development of this work, along with Elena Pannarale
for all her support.


\bibliographystyle{apsrev4-1-noeprint}
\bibliography{PhenoMixedSpinGWs_l}

\end{document}